\documentclass[conference]{IEEEtran}
\IEEEoverridecommandlockouts
\usepackage{cite}
\usepackage{booktabs}
\usepackage{threeparttable}
\usepackage{diagbox}
\usepackage{hhline}
\usepackage{float}
\usepackage{multirow}
\usepackage{makecell}
\usepackage[numbers,sort&compress]{natbib}
\usepackage{amsmath,amssymb,amsfonts}
\usepackage{algorithmic}
\usepackage{graphicx}
\usepackage{textcomp}
\usepackage{subfigure}
\usepackage{xcolor}
\usepackage{svg}
\usepackage{bm}
\usepackage[ruled]{algorithm2e} 
\usepackage{amsmath}
\usepackage{url}
\def\BibTeX{{\rm B\kern-.05em{\sc i\kern-.025em b}\kern-.08em
    T\kern-.1667em\lower.7ex\hbox{E}\kern-.125emX}}
\begin{document}
\bibliographystyle{IEEEtranN}
\title{ICStega: Image Captioning-based Semantically Controllable Linguistic Steganography}
\author{Xilong Wang, Yaofei Wang*, Kejiang Chen, Jinyang Ding, Weiming Zhang*, and Nenghai Yu
\thanks{This work was supported in part by the Natural Science Foundation of China under Grant 62102386, 62002334, 62072421, and 62121002, and by Anhui Initiative in Quantum Information Technologies under Grant AHY150400.

X. Wang, K. Chen, J. Ding, W. Zhang and N. Yu are with the School of Cyber Science and Technology, University of Science and Technology of China (E-mail: wxl20010508@mail.ustc.edu.cn, chenkj@ustc.edu.cn, source@mail.ustc.edu.cn, zhangwm@ustc.edu.cn, ynh@ustc.edu.cn).

Y. Wang is with the School of Computer Science and Information Engineering, Hefei University of Technology (E-mail: wyf@hfut.edu.cn).

* Corresponding author.}}
\maketitle

\begin{abstract}
Nowadays, social media has become the preferred communication platform for web users but brought security threats. 
Linguistic steganography hides secret data into text and sends it to the intended recipient to realize covert communication. Compared to edit-based linguistic steganography, generation-based approaches largely improve the payload capacity. However, existing methods can only generate stego text alone. Another common behavior in social media is sending semantically related image-text pairs. 
In this paper, we put forward a novel image captioning-based stegosystem, where the secret messages are embedded into the generated captions. Thus, the semantics of the stego text can be controlled and the secret data can be transmitted by sending semantically related image-text pairs.
To balance the conflict between payload capacity and semantic preservation, we proposed a new sampling method called Two-Parameter Semantic Control Sampling to cutoff low-probability words. Experimental results have shown that our method can control diversity, payload capacity, security, and semantic accuracy at the same time.
\end{abstract}


\section{Introduction}
\label{sec1}
Steganography is the science of hiding secret message in inconspicuous objects.
Among social media such as Twitter and Meta, images and text are two common formats for sharing. Although image steganography \cite{9414723,9414055,9054486} can carry plenty of messages, it cannot effectively resist the compression of social media. Linguistic steganography enjoys the advantage of high robustness during data transmission, which better ensures the success of covert communication in social media.

\par
\begin{figure}[t]
	\centering  
	\includegraphics[width =0.7\linewidth]{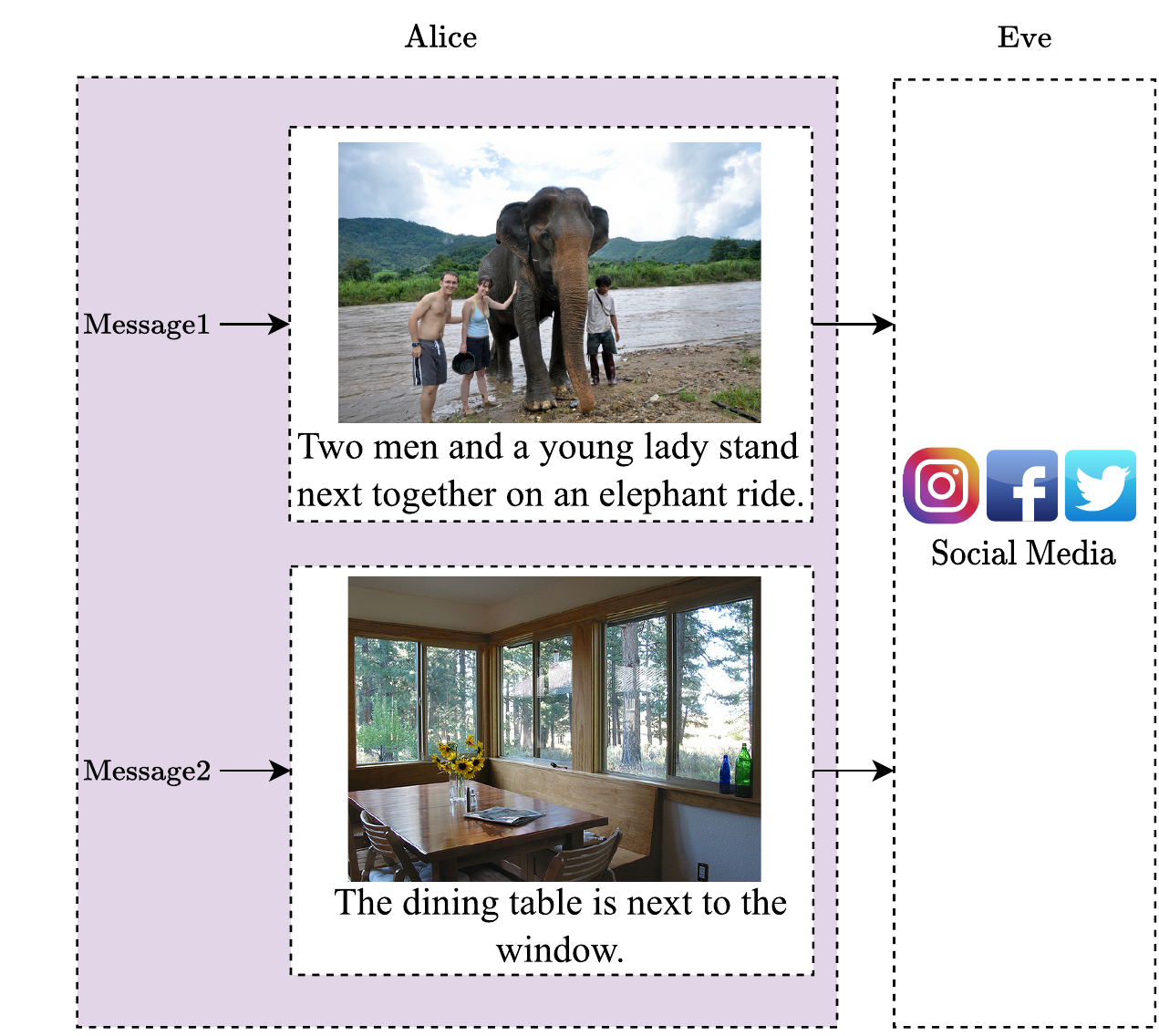}
	\caption{The application scenario of our steganography—posting stego text with a semantically related image, making the behavior more secure.}
	\label{fig1}
\end{figure}

Generally, linguistic steganography can be divided into two categories: edit-based \cite{ueoka2021frustratingly} and generation-based \cite{yang2018rnn,yang2020vae}. Edit-based approaches work by synonym substitution, thus achieving a certain semantic accuracy. However, restricted by little redundancy, they are inefficient and outperformed by generation-based approaches.
For generation-based linguistic steganography, recently, with advances in machine learning, statistical language models \cite{yang2018rnn,yang2020vae} that are close to natural language have been built. They have largely improved the content security, i.e., the indistinguishability between stego text and natural text. However, it is worth noting that most of the exiting generation-based linguistic steganography methods do not take into account image semantics, as sending a text with a semantically related image is a popular behavior in social media as shown in Figure \ref{fig1}. Furthermore, despite the work by Li et al. \cite{li2021topic}, most generation-based steganography cannot control the semantics of the stego text and thus cannot satisfy semantically controllable application scenarios. To solve the problem of semantic controllability, we apply image captioning to realize this idea.

\par
\par
Image captioning \cite{vinyals2015show,mokady2021clipcap,9747820} is the task of describing an image in words.
The words in the caption are predicted one by one according to the conditional probability distribution. In each step, image captioning models produce conditional probability for all words in the vocabulary, then, the next word is determined according to the conditional probability distribution. This feature allows us to perform generation-based linguistic steganography based on image captioning.
\par
In this paper, we propose a novel image captioning-based linguistic steganography (ICStega) where the secret messages are embedded into the generated captions. 
Similar to previous generation-based approaches, in each iteration, we build a candidate pool (CP) and sample words in CP.
In this process, previous works seek to make the stego text and natural language statistically indistinguishable. Hence, they need to create diversity and randomness for stego text. To this end, they fix the size of CP to include words with various probabilities. Unlike previous works, ICStega is based on image captioning. Hence, in addition to diversity and randomness, the generated captions need to be semantically relevant to the images. 
Therefore, we need to cutoff words that are semantically irrelevant. Hence, we put forward a new method called Two-Parameter Semantic Control Sampling, so as to cutoff the semantically irrelevant words.
\par
The main contributions are as follows:
 (a) We created a new application scenario for linguistic steganography where the secret messages are transmitted by sending the image-stego text pairs to social media.
(b) the stego text of our method is semantically controllable, which can describe images in words.
(c) we proposed an optimized strategy to build the candidate pool, so as to improve the semantic accuracy and further control the semantic expression of stego text. 

\section{Methods}\label{sec3}


\subsection{Overview}
\begin{figure}[t]
  \centering
  \includegraphics[width =\linewidth]{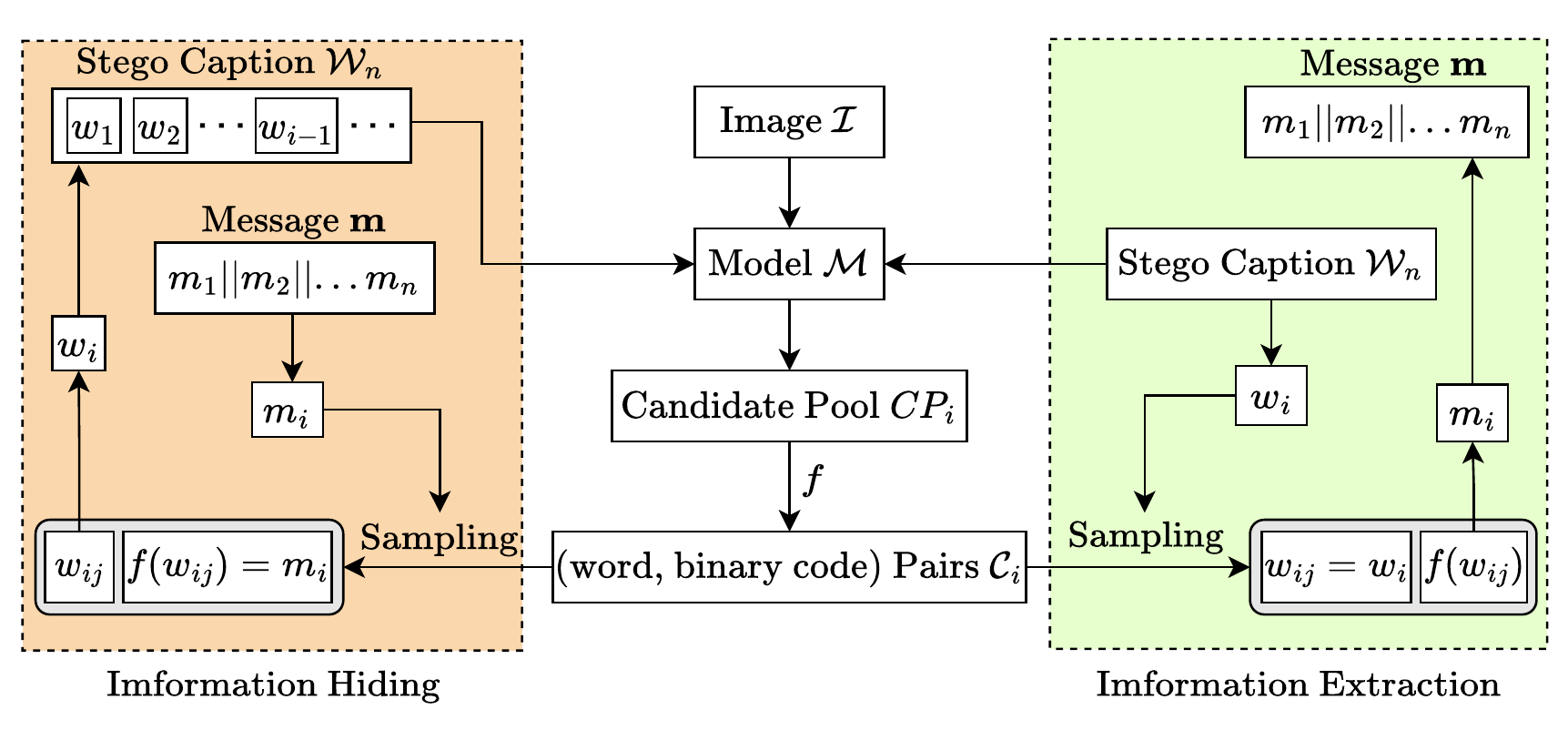}
  \caption{ The process of information hiding and extraction.}
  \label{fig3}
\end{figure}

The overview of image captioning-based linguistic steganography (ICStega) is shown in Fig.~\ref{fig3}. In the information hiding process, based on an uploaded image $\mathcal{I}$, the image captioning model $\mathcal{M}$ generates the conditional probability of words in the vocabulary. Next, a candidate pool (CP) containing words with the largest probabilities is built according to the conditional probability distribution. In this process, we propose a new method called Two-Parameter Semantic Control Sampling to cutoff words with low probabilities. 
Then, words in CP were assigned binary codes by a certain encoding strategy and sampled to form the stego caption carrying the secret message $\bm m$.

\subsection{Information Hiding Algorithm}


Let $V=\{v_1,v_2,...,v_l\}$ represent the vocabulary. For a caption $\mathcal{W}$ with length $n$, it can be represented as $\mathcal{W}_n=\{w_1,w_2,...,w_n\},w_i \in V$. As shown in \textbf{Algorithm 1}, based on previous output words $\mathcal{W}_{i-1} = \{w_1,w_2,...,w_{i-1}\}$, in the $i$-th step, we can attain $\mathcal{V}_i$—the conditional probability of every word in vocabulary $V$ by model $\mathcal{M}$. Then according to certain strategy, we can build a candidate pool $CP_i$ based on $\mathcal{V}_i$. Afterwards, we can encode each word $w_{ij}$ in the CP based on its conditional probability $p_{{ij}}$ by encoding strategy $f$ such as Huffman coding (HC) \cite{yang2018rnn}. Then we can choose the output word based on the secret message $\bm m$. How to build the candidate pool is the key issue to consider in this paper.
\par
If we do not consider the semantic association between text and image, it is better to choose the Top-K sampling as the previous generation-based steganography method did. 
However, as will be shown in Sec. \ref{sec4d}, using Top-K sampling would make the text and image unrelated. Hencce, in the next subsection, we introduce a new sampling method based on Top-K that preserves the advantages of Top-K and better maintains the semantics.


\begin{algorithm}[htbp] 
	\caption{Information Hiding Algorithm}
	\LinesNumbered 
	\KwIn{\\
	    Secret message $\bm m \sim Unif(\{0,1\}^L)$ \\
	    Image captioning model $\mathcal M$\\
	    Vocabulary $V$\\
	    Image $\mathcal I$\\
	    Encoding strategy $f$\\
	    Parameters $T_{A},T_{R}$\\
	    Two-Parameter Semantic Control Sampling $\mathcal{E}$
	    }
	\KwOut{\\stego text $ \mathcal{W}_n=\{w_1,w_2,...,w_n\},w_i \in V$}
	\While{not the end of $\bm m$}{
		$\mathcal{V}_i= \mathcal M(\mathcal I,\mathcal{W}_{i-1})$ \;
		$CP_i = \mathcal{E}(\mathcal{V}_i,T_{A},T_{R})$\;
		$\mathcal{C}_i = f({CP}_i)$\;
		\ForEach{$(w_{ij},f(w_{ij})) \in \mathcal{C}_i$}{
		    \If{$f(w_{ij})$ is a substring of $m$}{
		        $m_i=f(w_{ij})$\;
		        $w_i=w_{ij}$\;
		    }
		}
	}
	\textbf{return} $\mathcal{W}_n$
\end{algorithm}
\subsection{Two-Parameter Semantic Control Sampling}
\par

\begin{figure}[t]
  \centering
  \resizebox{0.8\linewidth}{!}{
  \includegraphics[width =0.9\linewidth]{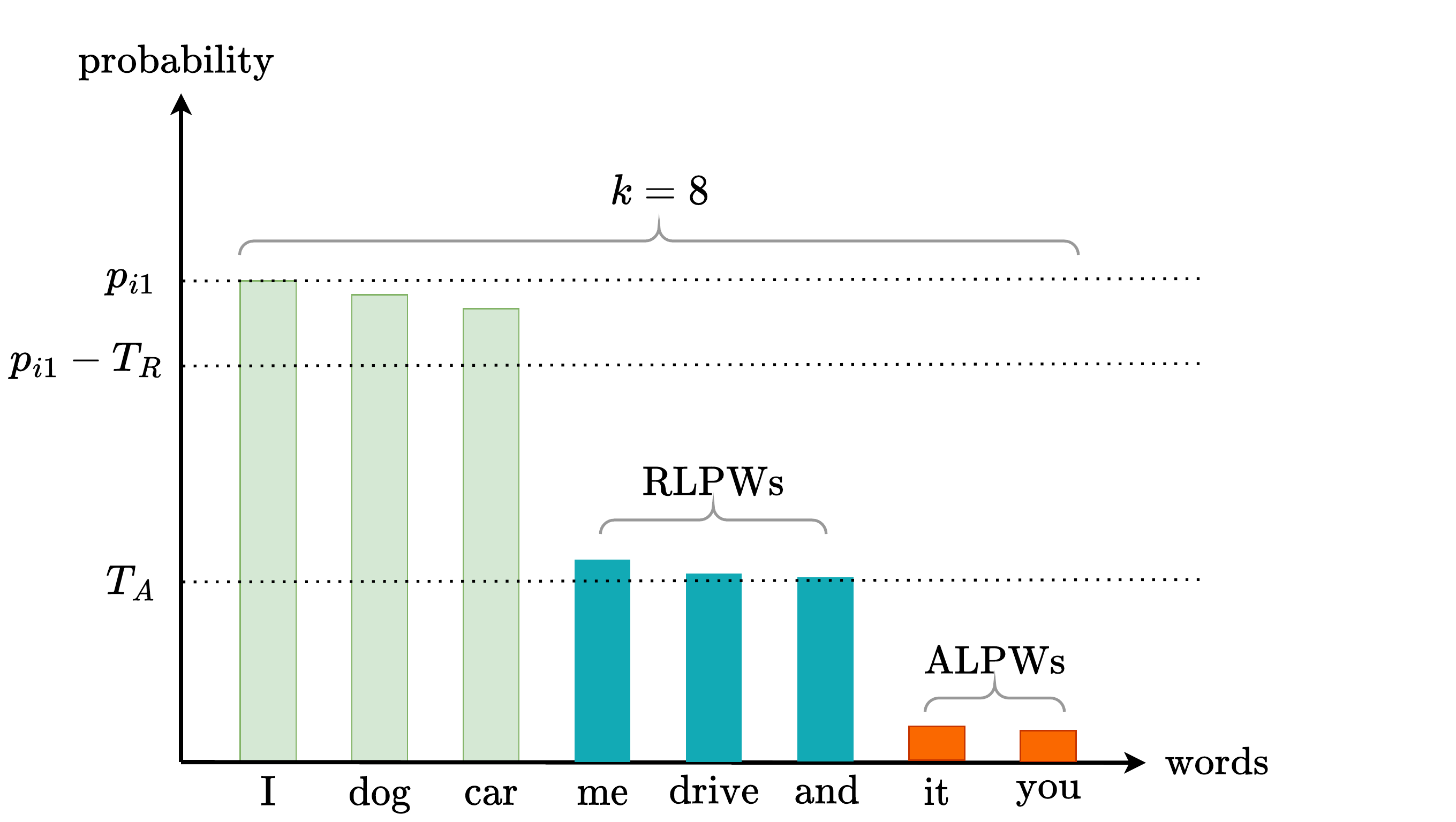}
  }
  \caption{ Different strategies to build candidate pool. When $k$ is fixed, both \textbf{ALPW}s and \textbf{RLPW}s cannot be cutoff. When we only implement $T_{A}$, it can only cutoff the \textbf{ALPW}s. If we apply $T_{A}$ and $T_{R}$, both \textbf{ALPW}s and \textbf{RLPW}s will be cutoff.}
  \label{fig4}
\end{figure}
Generally, low-probability words can be classified into two types: relatively low-probability words (\textbf{RLPW}s) and absolutely low-probability words (\textbf{ALPW}s). As shown in Fig.~\ref{fig4}, when the size of the candidate pool $k$ is fixed to $8$, sometimes the \textbf{ALPW}s will be included in the candidate pool. To cutoff them, we propose a new parameter $T_{A}$, and the word $w_{ij}$ will be included in the candidate pool only if $p_{ij} > T_{A}$. Nevertheless, we still cannot obtain an ideal candidate pool with a fine probability distribution. As we can see in Fig.~\ref{fig4}, even with the implementation of $T_{A}$, probabilities still vary widely from word to word. The probabilities of words \{“dog”, “car”\} are very close to the word with the largest probability \{“I”\}, and they are significantly higher than probabilities of \{“me”, “drive”, “and”\}. Hence, \{“I”, “dog”, “car”\} can all be regarded as “suitable words”, while \{“me”, “drive”, “and”\} can be classified as “unsuitable words”, i.e, \textbf{RLPW}s. In general, words whose possibilities are relatively close to the largest one $p_{i1}$ should be regarded as “suitable words” and included in the candidate pool, and we should cutoff not only the \textbf{ALPW}s but also the \textbf{RLPW}s. In addition, the \textbf{RLPW}s cannot be swiped by simply increasing $T_{A}$—if $T_{A}$ is too high, the “suitable words” might also be cutoff. To thoroughly filter out those \textbf{RLPW}s and preserve the “suitable words”, we introduce another parameter $T_{R}$. It is used for judging the 
distance between $p_{ij}$ and the largest probability $p_{i1}$. If $p_{ij}$ and $p_{i1}$ are very close, $w_{ij}$ will be regarded as “suitable words”. To be detailed, $w_{ij}$ will be preserved only if $p_{i1}-p_{ij}<T_{R}$. Considering $T_{A}$ and $T_{R}$ together, $w_{ij}$ will be included in $CP_i$ only if $p_{ij}>\max(T_{A}, p_{i1}-T_{R})$. With $T_{A}$ and $T_{R}$, we can cutoff \textbf{ALPW}s and \textbf{RLPW}s without losing “suitable words”. 

\section{Experiments and Analysis}\label{sec4}
In this section, we first tested the payload capacity and imperceptibility of ICStega and
then selected the suitable $T_{A}$ and $T_{R}$ for further evaluation. 
Then, we tested the security performance and image captioning performance. For comparison, we employed the typical generate-based steganography RNN-Stega \cite{yang2018rnn} and edit-based steganography MLStega \cite{ueoka2021frustratingly}.
For ICStega, for the image captioning model $\mathcal{M}$, we trained the ClipCap model \cite{mokady2021clipcap} based on the MS COCO dataset \cite{lin2014microsoft,chen2015microsoft}. For each $T_{A}$ and $T_{R}$, we generated 1,000 stego captions by Huffman coding (HC) based on 1,000 images. For RNN-Stega \cite{yang2018rnn}, we trained it based on the Twitter \cite{go2009twitter} dataset and encoded the top 2, 4, 8, ..., 64 likely words by HC. 
For MLStega \cite{ueoka2021frustratingly}, we randomly sampled 1,000 captions from the MS COCO validation set and embedded secret messages by editing these captions under different intervals $\delta$ and thresholds $p$.

\subsection{Parameter Optimization}
\par
Here we first compare the payload capacity with unit of bits per word (bpw).
Experimental results are shown in TABLE \ref{tab1}, from which we can draw the following conclusions. First, the payload capacity of ICStega is not lower than RNN under certain conditions. 
Second, ICStega has better payload capacity than MLStega, in which secret messages are only embedded at specific locations in sentences.

\begin{table}[t]
\centering
\caption{bpw for different settings and methods.}
\resizebox{0.9\linewidth}{!}{
\begin{tabular}{@{}|ccccccc|@{}}

\hline
 \multicolumn{7}{|c|}{ICStega}                           \\
\hline
                               \multicolumn{1}{|c|}{\diagbox{$T_{R}$}{bpw}{$T_{A}$}}      & 0     & 0.001  & 0.005  & 0.01   & \multicolumn{1}{c}{0.05}   & 0.1   \\
\hline
                               \multicolumn{1}{|c|}{0.1}  & 0.290 & 0.287 & 0.283 & 0.284 & \multicolumn{1}{c}{0.258} & \multicolumn{1}{c|}{0.219} \\
                              \multicolumn{1}{|c|}{0.2}  & {0.706} & 0.618& 0.602 & {0.582} & \multicolumn{1}{c}{0.486} & \multicolumn{1}{c|}{0.354} \\
                               \multicolumn{1}{|c|}{0.3}  & {1.092} & {0.935} & {0.866} & {0.811} & \multicolumn{1}{c}{\textcolor{red}{\textcolor{red}{\textbf{0.627}}}} & \multicolumn{1}{c|}{0.455} \\
                              \multicolumn{1}{|c|}{0.4}  & {1.333} & {1.105} & {1.015} & {0.981} & \multicolumn{1}{c}{{0.688}} & \multicolumn{1}{c|}{{0.522}} \\
                              
                              \multicolumn{1}{|c|}{0.6}  & {1.575} & {1.370} & {1.239} & {1.136} & \multicolumn{1}{c}{{0.755}} & \multicolumn{1}{c|}{{0.529}} \\
                               \multicolumn{1}{|c|}{0.8}  & {2.391} & {1.979} & {1.822} & {1.558} & \multicolumn{1}{c}{{1.050}} & \multicolumn{1}{c|}{0.618} \\
                               \multicolumn{1}{|c|}{1}    & {3.145} & {2.576} & {2.133} & {1.897} & \multicolumn{1}{c}{{1.107}} & \multicolumn{1}{c|}{0.607} \\
                              
\hline
                              \multicolumn{7}{|c|}{RNN-Stega \cite{yang2018rnn}}                                                                  \\
\hline
                               \multicolumn{1}{|c|}{$top\ k$} & 2   & 4   & 8   & 16   & 32                        & 64                       \\
\hline
                               \multicolumn{1}{|c|}{bpw}   & 1.000     & 1.880    &2.649 & 3.375     & 3.972     & 4.592                           \\
\hline
                              \multicolumn{7}{|c|}{MLStega \cite{ueoka2021frustratingly}}\\
\hline
 \multicolumn{1}{|c|}{$\delta$}   & \multicolumn{2}{c|}{2}                              & \multicolumn{2}{c|}{3}                             & \multicolumn{2}{c|}{4}                                                   \\
\hline
 \multicolumn{1}{|c|}{$p$}    & 0.01  & \multicolumn{1}{c|}{0.05}               & 0.01 & \multicolumn{1}{c|}{0.05}   & 0.01 & 0.05                       \\
\hline
 \multicolumn{1}{|c|}{bpw}    & 0.624   & \multicolumn{1}{c|}{0.203}    & 0.369         & \multicolumn{1}{c|}{0.134}     & 0.300      & \multicolumn{1}{c|}{0.113}                          \\

\hline

\end{tabular}
}
\label{tab1}
\end{table}

\par
One of the central goals of steganography is to pursue ``imperceptibility", i.e., the language model used for steganography should generate stego sentences with higher quality while embedding enough confidential messages. In natural language processing, perplexity ($ppl$) is a measurement used to evaluate the quality of a generated sentence.
It is defined as the average per-word log-probability on the test texts: 
\begin{equation}
    \begin{aligned}
        ppl&=2^{-\frac{1}{n} \log p(\mathcal{W}_n)}=2^{-\frac{1}{n} \sum ^n_{j=1} \log p(w_j|w_1,...,w_{j-1})}
    \end{aligned}\label{eq10}
\end{equation}
The lower $ppl$ is, the higher the quality of the stego sentence. In this paper, we calculate the mean $ppl$ of ICStega under different parameters as well as that of RNN-Stega. Experimental results are recorded in TABLE \ref{tab2}.
\begin{table}[]
\centering
\caption{The Mean $ppl$ of Each Method}
\resizebox{0.9\linewidth}{!}{
\begin{tabular}{@{}|ccccccc|@{}}
\hline
 \multicolumn{7}{|c|}{ICStega}                           \\
\hline
                               \multicolumn{1}{|c|}{\diagbox{$T_{R}$}{$ppl$}{$T_{A}$}}      & 0     & 0.001  & 0.005  & 0.01   & \multicolumn{1}{c}{0.05}   & 0.1   \\
\hline
                                \multicolumn{1}{|c|}{0.1}  & {2.056}& {2.047} & {2.049} & {2.037} & {2.005} & {{1.996}}  \\
                                \multicolumn{1}{|c|}{0.2}  &{2.408}& {2.377}& {2.253} & {2.234} & {2.146} & {2.072} \\
                                \multicolumn{1}{|c|}{0.3}  &3.094& 2.781 & 2.566 & 2.510 & \textcolor{red}{\textbf{2.270}}  & {2.145} \\
                                \multicolumn{1}{|c|}{0.4}  &3.608& 3.222 & 2.935 & 2.725 & 2.400  & {2.206} \\
                               \multicolumn{1}{|c|}{0.6}  &5.071& 4.156 & 3.472 & 3.293 & 2.659  & {2.330} \\
                               \multicolumn{1}{|c|}{0.8} &7.685 & 5.510 & 4.645 & 4.164 & 3.126  & 2.451 \\
                               \multicolumn{1}{|c|}{1}    & 14.525& 8.901  &  6.393&5.362  & 3.207  & 2.429 \\

\hline
                                \multicolumn{7}{|c|}{RNN-Stega \cite{yang2018rnn}}                                                                  \\
\hline
                               \multicolumn{1}{|c|}{$top\ k$} & 2   & 4   & 8   & 16   & 32                        & 64                       \\
\hline
                                \multicolumn{1}{|c|}{$ppl$}  & 19.903     & 23.326     & 29.988     & 38.134     & 46.518                          & 57.560                           \\

\hline

\end{tabular}
}
\label{tab2}
\end{table}
According to TABLE \ref{tab1} and \ref{tab2}, under a similar payload, ICStega has lower $ppl$ than RNN-Stega, which means that it has higher imperceptibility. For example, for ICStega, when $T_{A}=0.005$ and $T_{R}=0.4$, $ppl$ is 2.935 (with 1.015 bpw), and for RNN-Stega, when $top \ k=2$, $ppl$ is 19.903 (with 1.000 bpw). 
\par
To control the trade-off between imperceptibility and payload capacity, we need to adjust $T_{A},\ T_{R}$ and select the optimal parameters. For our method, we pick situations whose payload $>$0.620 bpw and $ppl$ $<$ 2.400. The results corresponding to the optimal parameters are shown in red font in TABLE \ref{tab1} and TABLE \ref{tab2}: we pick $T_{A}=0.05,\ T_{R}=0.3$ for ICStega.
\subsection{Security Performance}
In this part, we test the security performance of our method. we employ a steganalysis model TS-FCN \cite{yang2019fast} to detect stego sentences generated under different $T_{A}$ and $T_{R}$. 
We implement $Accuracy$ ($Acc$) as the evaluation indicator of steganalysis model, which describes the accuracy of a steganalysis model to classify stego text and real text. 
%
The closer $Acc$ is to 0.5, the better the steganography model is.
\begin{table}[]
\centering
\caption{Security Performance of Each Proposed Method under Different Conditions and Encoding Strategies}
\resizebox{0.88\linewidth}{!}{
\begin{tabular}{@{}|ccccccc|@{}}
\hline
    \multicolumn{7}{|c|}{ICStega}                           \\
\hline
                              \multicolumn{1}{|c|}{\diagbox{$T_{R}$}{$Acc$}{$T_{A}$}}      & 0     & 0.001  & 0.005  & 0.01   & \multicolumn{1}{c}{0.05}   & 0.1   \\
\hline
                               \multicolumn{1}{|c|}{0.1}  & 0.743 &0.744 &0.741 & 0.739 & 0.745 & 0.761  \\
                               \multicolumn{1}{|c|}{0.2}  & 0.644 &0.670 &0.670 & 0.668 & 0.687 & 0.727 \\
                               \multicolumn{1}{|c|}{0.3}  & 0.591 &0.608 &0.611 & 0.613 & \textcolor{red}{\textbf{0.650}} & 0.693 \\
                               \multicolumn{1}{|c|}{0.4}  & 0.558 &0.570 &0.591 & 0.583 & 0.625 & 0.673 \\
                               \multicolumn{1}{|c|}{0.6}  & 0.509 &0.506 &0.525 & 0.527 & 0.574 & 0.619  \\
                               \multicolumn{1}{|c|}{0.8}  & 0.531 &0.518 &0.550 & 0.534 & 0.580 & 0.612 \\
                               \multicolumn{1}{|c|}{1}    & 0.598 &0.608 &0.574 & 0.587 & 0.561 & 0.628 \\
                              
\hline

                                 \multicolumn{7}{|c|}{RNN-Stega \cite{yang2018rnn}}                                                                  \\
\hline
                                 $top\ k$ & 2   & 4   & 8   & 16   & 32                        & 64                       \\
\hline
                                 $Acc$  & 0.802     & 0.767     & 0.748     & 0.706     & 0.683                          & 0.686                           \\
\hline

                               \multicolumn{7}{|c|}{MLStega \cite{ueoka2021frustratingly}}\\
\hline
 \multicolumn{1}{|c|}{$\delta$}   & \multicolumn{2}{c|}{2}                              & \multicolumn{2}{c|}{3}                             & \multicolumn{2}{c|}{4}                                                   \\
\hline
\multicolumn{1}{|c|}{$p$}   &  0.01   & \multicolumn{1}{c|}{0.05}    & 0.01  & \multicolumn{1}{c|}{0.05}  &  0.01 & 0.05                       \\
\hline
 \multicolumn{1}{|c|}{$Acc$}                                                   & 0.633         & \multicolumn{1}{c|}{0.571}         & 0.541     & \multicolumn{1}{c|}{0.552}                              & 0.549         & \multicolumn{1}{c|}{0.518}                          \\               
\hline

\end{tabular}
}
\label{tab3}
\end{table}
\par
The experimental results are recorded in TABLE \ref{tab3}. Based on TABLE \ref{tab1} and TABLE \ref{tab3}, we can draw the following conclusions. First, under a similar payload, ICStega has better security performance than RNN-Stega. For example, for ICStega, when $T_{A}=0.005$ and $T_{R}=0.4$, the $Acc$ is 0.591 (with 1.015 bpw), and for RNN-Stega, when $top\ k=2$, the $Acc$ is 0.802 (with 1.000 bpw). Second, 
compared to MLStega \cite{ueoka2021frustratingly} which has advantages in content security, our method shows a greater performance on payload capacity without the loss of content security: for example, for ICStega, when $T_{A}=0.001,\ T_{R}=0.4$, $Acc$ is 0.570 with 1.105 bpw, and for MLStega, $Acc$ is 0.571 with 0.203 bpw.
\subsection{Image Captioning Evaluation}
\label{sec4d}
In this paper, we propose an image captioning-based steganography approach,  
so we need to consider the accuracy of captions, i.e., how relevant captions and images are. To this end, we use two evaluation metrics BLEU$_4$ \cite{papineni2002bleu} and METEOR \cite{denkowski2014meteor}. In addition, we utilize LSA and Self-CIDEr \cite{wang2019describing} to measure the diversity of captions, i.e., describing a single image with different captions.



\par
We randomly choose 1,000 images and generate 10 captions for each image under the optimal parameters ($T_{A} = 0.05,\ T_{R}=0.3$). For comparison, we generate the same amount of captions by Top-K sampling without embedding secret messages. The results are shown in TABLE \ref{tab4}. 
\begin{table}[b]
\centering
\caption{Image Captioning Performance of Each Situation}

\begin{tabular}{|c|c|c c | c c|}

\hline
 \multirowcell{7}{Captions\\with\\Top-K} & 
\multirow{1}{*}{$top\ k$} &
\multirow{1}{*}{BLEU$_4$$\uparrow$} &
\multirow{1}{*}{METEOR$\uparrow$} &
\multirow{1}{*}{LSA$\uparrow$} &
\multirow{1}{*}{Self-CIDEr$\uparrow$}\\
\cline{2-6}
&2 &	0.320 &	0.278 &0.506 &0.721
\\
 &4 & 0.251	&0.257 &0.596 &0.833
\\
 &8 	&0.215	&0.243 &0.627 &0.871
\\
 &16 &0.204& 0.233&0.661 &0.891
\\
 &32 	&0.187	&0.228 & 0.657 &0.897\\
 &64  &0.177  & 0.224 &0.672 &0.905\\
\hline
\multicolumn{2}{|c|}{ICStega} &0.312	&0.271 &0.517  & 0.717\\
 
\hline
\end{tabular}

\label{tab4}
\end{table}
For captions generated by Top-K sampling, varying $top\ k$, the accuracy and diversity show opposite tendencies: the accuracy decreases but diversity increases as $top\ k$ grows. This is because when $top\ k$ increases, more words with lower probability are included. And the previous generation-based steganography such that RNN-Stega \cite{yang2018rnn} often choose a bigger $top\ k$ for a higher payload as shown in TABLE \ref{tab1} and that lead the text and image unrelated. However, ICStega does not degrade the accuracy while reaching a certain degree of diversity, thus finding a balance between these two factors with the benefit of Two-Parameter Semantic Control Sampling. 
\par
Moreover, we also employ other approaches RNN-Stega \cite{yang2018rnn} and MLStega \cite{ueoka2021frustratingly} for comparison. We randomly choose an image-ground truth caption pair from the MS COCO \cite{lin2014microsoft} validation set and take the first several words as context. Then, based on the context, we apply RNN-Stega to complete the caption. For MLStega, we embed secret messages by editing the ground truth caption
. Examples are shown in TABLE \ref{fig5}. For ICStega, 
the semantics of the generated captions are still highly relevant to the image. However, for RNN-Stega, it is obvious that the produced captions are digressive. For MLStega, although it will only edit a small number of words in the sentence, the key information of the caption may be changed.
For example, for the second case, it converts “table” to
“map”, but there is no map in the image.
\begin{table}[htbp]
    \centering
    \caption{Captions Processed by Different Methods.}
\resizebox{\linewidth}{!}{
    \begin{tabular}{c| c |c}
    \hline
        Image &     \begin{minipage}[c]{0.25\linewidth} 
                \centerline{\includegraphics[width=3.6cm]{COCO_val2014_000000000136.jpg}}
                \end{minipage}  &
                \begin{minipage}[c]{0.25\linewidth} 
                \centerline{\includegraphics[width=3.6cm]{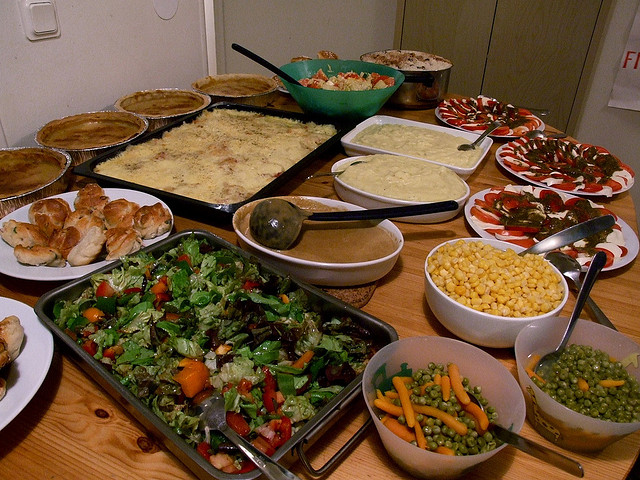}}
                \end{minipage}
                
        \\
        \hline
        \multirow{2}{*}{Ground Truth} & Two giraffes standing next &A variety of food in dishes\\&to each other in a room.&is displayed on a table.\\
        \hline
        \multirow{2}{*}{MLStega \cite{ueoka2021frustratingly}} & \textcolor{cyan}{Three} giraffes standing \textcolor{cyan}{close}& A \textcolor{cyan}{piece} of food in \textcolor{cyan}{rows} \\
        & to each other in a room.&is displayed on a \textcolor{cyan}{map.}\\
        \hline
        \specialrule{0em}{1.25pt}{1.25pt}
        \hline
        \multirow{2}{*}{ICStega} & Two giraffes stand in & A table with several
 \\
         & {a room with a metal fence.} & different types of food on it.\\
        \hline
        \specialrule{0em}{1.25pt}{1.25pt}
        \hline
        Context & Two giraffes standing &A variety of food in dishes\\
        \hline
        \multirow{2}{*}{RNN-Stega \cite{yang2018rnn}} & Two giraffes standing& A variety of food in dishes \\
         &  \textcolor[rgb]{1,0,1}{in the sun and the sun is out} & \textcolor[rgb]{1,0,1}{a few hours}\\
        \hline
    \end{tabular}
}
    \label{fig5}
\end{table}

\section{Conclusion}\label{sec5}
In this paper, we put forward a novel image captioning-based linguistic steganography (ICStega), where the semantics of the stego text can be controlled. Unlike previous generation-based methods, as an image captioning task, we need to pay special attention to the semantic accuracy of stego captions. Therefore, we also put forward a new method called Two-Parameter Semantic Control Sampling to cutoff words with low probabilities. We experimentally compare the effects of different parameters on payload, imperceptibility, and choose the appropriate parameters for steganography. We show through metrics and examples that our approach not only has better payload capacity, imperceptibility, and security performance but also can properly describe the content of the image, which satisfies the need of sending the semantically related image-text pairs in social media. In future work, we will further enhance the security, semantic accuracy, and certain robustness without losing payload capacity.
\bibliography{IEEEabrv,references}
\end{document}